# Recent Advances in Microfluidics and Bioelectronics for Three-Dimensional Organoid Interfaces

Caroline Ferguson[a,b], Yan Li[c], Yi Zhang[a,*], Xueju Wang[b,*]

[a] Department of Biomedical Engineering, College of Engineering, University of Connecticut, 260 Glenbrook Rd., Storrs, CT 06269, USA

[b] Department of Materials Science and Engineering, College of Engineering, University of Connecticut, 25 King Hill Rd., Storrs, CT 06269, USA

[c] Department of Chemical and Biomedical Engineering, FAMU-FSU College of Engineering, Florida State University, 2525 Pottsdamer St., Tallahassee, FL 32310, USA

## Abstract:

Organoids offer a promising alternative in biomedical research and clinical medicine, with better feature recapitulation than 2D cultures. They also have more consistent responses with clinical results when compared to animal models. However, major challenges exist in the longevity of culture, the reproducibility of organoid properties, and the development of non-disruptive monitoring methods. Recent advances in materials and microfabrication methods, such as 3D printing and compressive buckling, have enabled three-dimensional (3D) interfaces of microfluidics and bioelectronics to manipulate and monitor these biological models in exciting ways. These advanced systems have great potential for applications in drug delivery, personalized medicine, and disease modelling. We conclude with important future considerations to generate longevity using further technological development in organoid and spheroid models.

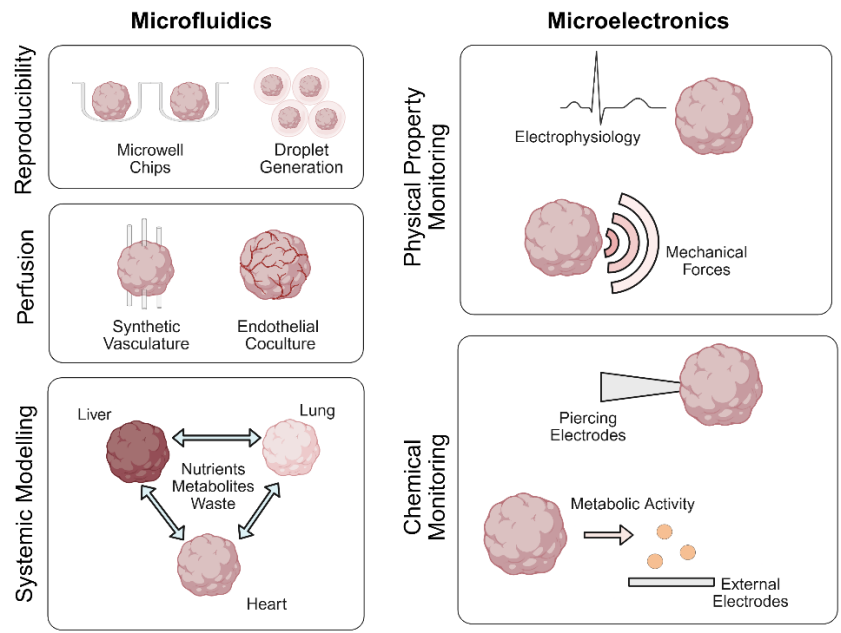

*Keywords:* Organoids, Organ-on-a-Chip, 3D Flexible Electronics, 3D Microfluidics

---

# 1. Organoid Culture Overview

The development of *in vitro* models to replicate complex cell assemblies of the human body has been a longstanding goal in biotechnology. While two-dimensional (2D) cultures have been standard in a laboratory setting, three-dimensional (3D) multicellular structures more accurately represent the interactions and processes occurring within the human body, making them more reliable models for widespread analysis [1,2]. Advantages of 3D *in vitro* models over planar models include insight into tissue and disease development, advancing understanding of cell-cell interactions within the extracellular matrix (ECM) to generate more effective biomaterials, and enhancing drug testing platform accuracy before clinical trials, thereby reducing reliance on animal models [3].

Recognizing these benefits, the Food and Drug Administration (FDA) has recently approved organoids as suitable proof-of-concept drug screening models before human clinical trials, improving the efficiency of drug development and transition into clinical trials. Recent legislation, including the FDA Modernization Act 2.0, supports this transition by approving alternative methods to evaluate drugs, including the use of organ-on-a-chip technology, organoid models, and predictive machine learning models [4]. These advancements underscore the importance of developing and refining high-fidelity organoid and spheroid models, positioning them as critical tools in the future of drug discovery and biotechnology.

## 1.1 Sources and Generation

The general structure of 3D biological model development is 1) selection of the source cell type, 2) generation of a multidimensional structure, 3) period of extended growth, and 4) evaluation. These models are described as organoids, spheroids, and assembloids based on the cell type and generation method. Most cells can form 3D structures when placed in the appropriate conditions, including immortalized cell lines, sourced primary cells, and induced pluripotent stem cells (iPSCs). Each cell type has different properties, like growth rate, accessibility, and resource requirements, that may make it an advantageous choice. Primary cells are challenging to source and often require customized protocols to thrive in laboratory settings [5]. Similarly, iPSCs require extensive protocols, needing substantial growth factors, trained personnel, and time to reach sufficient cell numbers that recapitulate *in vivo* niche conditions [6,7]. In contrast, immortalized cell lines are more accessible, less resource-intensive, and proliferate quickly; however, they lack the diversity of cell types and structural complexity found in other sources. The selection of cell types depends on balancing available resources and the desired model complexity. These considerations are essential for both defining the model as a spheroid, organoid, or assembloid and designing an appropriate generation process to replicate the desired spatial structures or dynamic cell interactions.

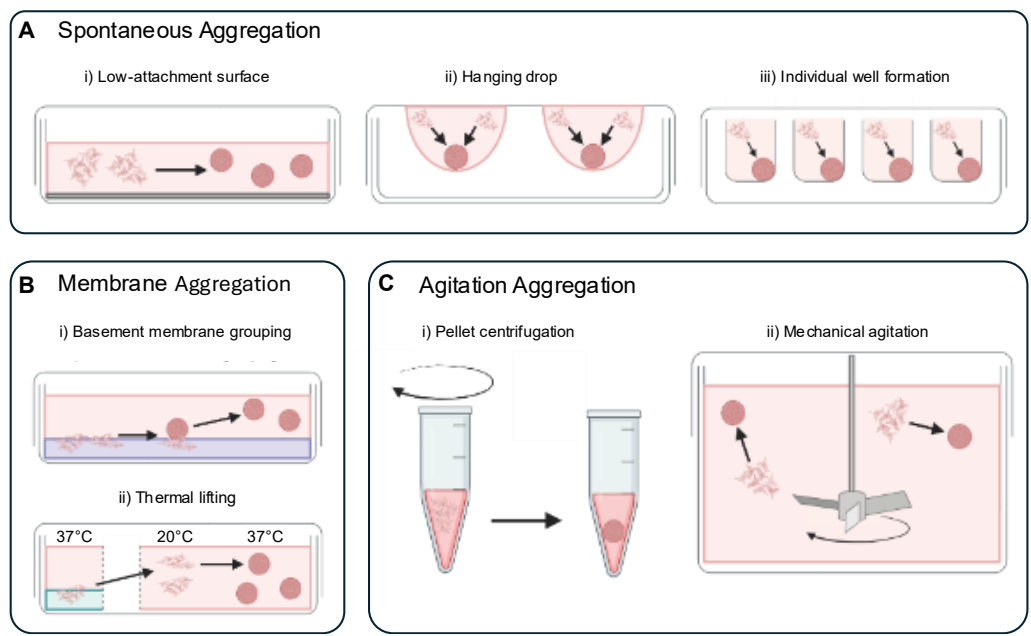

**Figure 1: Formation methods for spheroid models. A)** Methods for spontaneous generation of three-dimensional organoid structures including i) low attachment culture surfaces, ii) hanging drop, and iii) individual well formation of spheroids in a 96-well plate. **B)** Membrane-based aggregation methods using i) charged basement membranes or ii) thermal lifting processes. **C)** Agitation-based aggregation of spheroids based on i) initial centrifugation to form a pellet or ii) continuous mechanical agitation through a bioreactor. Created in BioRender.com.

3D structure formation typically begins after the introduction of relevant growth factors or genetic modifications that initiate specific developmental or disease states, leading the cells toward the desired differentiation. A common approach to achieving 3D growth is spontaneous aggregation, illustrated in Figure 1A, through i) culturing cells on a surface containing a low-attachment coating to promote clustering, ii) by leveraging gravity to encourage cell clustering at the bottom of a droplet, or iii) using small wells for increased interactions. Less passive approaches include membrane-based methods, as shown in Figure 1B, wherein a synthetic ECM scaffold is introduced to promote cell grouping [8,9] or utilizing the thermodynamic properties of a gel to lift cells upon dissolution. The ECM, generally Matrigel or a synthetic matrix, provides a layer with interspersed cells that support cell grouping into organoid formations and can have variable structures that limit reproducibility between batches [10]. Mechanical methods, represented in Figure 1C, include centrifugation into a cell pellet and bioreactor systems, where continuous stirring groups the cells into aggregates during culture. After formation, a period of growth continues to allow further properties and interactions to develop. Most critically, organoids and spheroids experience growth restriction due to the limited diffusion of oxygen and nutrients in the core region, resulting in constrained size and maturation.

As mentioned earlier, the classification of multicellular structures such as spheroids or organoids depends on their structural complexity. Nomenclature has been widely debated between spheroids and organoids, especially in subfields such as brain organoids, where a strong understanding of relevant terms for different parts of the brain being cultured becomes necessary to verify results between research groups [11]. Spheroids are generally simpler, composed of one or two cell types, and are commonly derived from immortalized cell lines. Historically, the term "spheroid" also refers to early-stage organoid development, whereas the term "organoid" indicates the development of more mature structures [12]. Assembloids represent a further step in complexity, combining multiple organoids or spheroids to generate further interactions, including an assembloid combining organoids that represent different specific regions of the brain. While our focus is primarily on 3D organoid interfaces, we also include spheroid interfaces in our discussion. As the breadth of organoid-based studies has expanded, analytical methods have also extended to fully capture, describe, and compare intricate multidimensional cellular properties.

### 1.2 Traditional Analysis Methods

Analytical methods are essential for establishing the identity and characteristics of cells, chemicals (like reactive oxygen species), and processes (like stem cell differentiation or cell maturation) that are key to forming organoid structures. These methods can be categorized according to their focus on compositional, structural, or genetic information, as illustrated in **Figure 2**. Given that these methods have been reviewed extensively elsewhere [1,13,14], we provide only a brief overview here.

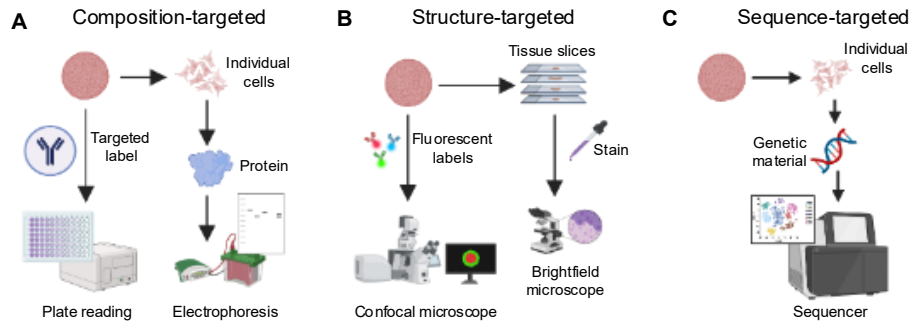

**Figure 2: Evaluation methods for analyzing spheroid structures. A)** Composition-targeted methods, including examples of targeted labeling and evaluation of protein content. **B)** Structure-targeted methods, including whole imaging with fluorescent labels or tissue sectioning for slide-based staining. **C)** Sequence-targeted evaluation, based on the dissociation of genetic material from individual cells for eventual DNA or RNA sequencing. Created in BioRender.com.

Composition-targeted methods utilize labelled molecules that bind to proteins released to the culture media by the cells or extracted from the cell structure, producing either colorimetric or fluorescent signals [13]. These assays are critical for assessing organoid function, as they can quantify metabolites or track the flux of signaling molecules such as calcium. Common assays include the 3-(4,5-dimethylthiazol-2-yl)-2,5-diphenyltetrazolium bromide (MTT) assay, which measures metabolic activity, and the Western Blot assay, which quantifies specific proteins, as shown in Figure 2A. Structure-targeted methods typically employ high-resolution imaging techniques to visualize organoids, either through label-free approaches such as bright-field microscopy, or by utilizing specific stains used in immunofluorescence or histology analyses [13]. Such techniques provide detailed spatial and temporal insights into the organoid structure, but often require specialized equipment or skills due to the complexity of analyzing 3D structures [14]. For example, specialization is required for processing, which can involve physical sectioning or digital reconstruction through stacked image slices, as illustrated in Figure 2B. Sequence-targeting analysis methods (Figure 2C) involve sequencing genetic materials extracted from either the bulk solution surrounding an organoid or from individual cells through techniques including quantitative polymerase chain reaction (q-PCR) or single-cell ribonucleic acid (RNA) sequencing

[14]. These techniques generate insights into the cell types and physiological expression post-differentiation but also require advanced equipment and expertise.

Despite their widespread use, traditional analytical methods face several limitations, particularly regarding culture longevity and spatiotemporal resolution. For example, many of these techniques necessitate removing organoids from their culture environment, thus halting further development and precluding longitudinal analysis of individual organoids [15]. Moreover, these methods are resource-intensive, requiring reagents, specialized equipment, and skilled personnel for processes like staining, advanced microscopy, and genetic sequencing [16]. The intrinsic heterogeneity of organoids—variability in size, structure, and cellular composition—also complicates reproducibility, often necessitating large sample sizes for statistically meaningful results [17]. Additionally, the laborious nature of organoid production and the associated data analysis pose further challenges.

Given these limitations, there is an urgent need for new tools and technologies that enable real-time, user-friendly, *in situ* monitoring of key biological processes, including proliferation, differentiation, and region-specific tissue patterning in functional organoid models. This review presents recent advances in integrating innovative microfluidics and bioelectronics with organoid systems, highlighting how these technologies overcome current analytical limitations. We also explore their convergence with emerging applications and propose strategies for building integrated 3D platforms to improve both the growth and characterization of organoids.

## 2. Microfluidic Organoid Interfaces

Current human organoids face limitations in structural maturity, functionality, size, and heterogeneity. They are largely attributed to the lack of stable, perfusable vascularization—an essential feature of native tissue that enables efficient nutrient and oxygen delivery, metabolic waste removal, and distribution of growth factors necessary for sustaining organoid viability, growth, and complexity[19,1,29]. For example, most organoids develop a necrotic core as the increased metabolic demands for cells located in the center cannot be met by passive diffusion of nutrition (effective distance of diffusion ≲~300 μm). In addition, tissue growth and identity rely on specific growth factors, including Wnt, Sonic Hedgehog, and Transformation Growth Factor β (TGF-β) [13], which play a vital role in regulating stem cell differentiation and enabling activities

(e.g., protein signaling) that coordinate the formation of complex organoid structures. Spatiotemporal control over growth factor distribution is essential to mimic the in vivo microenvironment, which is particularly relevant for applications in personalized medicine and drug development [13,18].

Achieving the level of precision to mimic the human body in terms of perfusion and multi-organ modeling is a complicated goal, especially considering the need for reproducible conditions to make statistical conclusions. Microfluidic technologies have been introduced to address challenges around three primary objectives: 1) enabling precise control over organoid formation for reproducible size and development, 2) improving continuous perfusion throughout organoid development to generate enhanced longevity of culture and maturity of organoid, and 3) developing systemic flow designs for more complex models to better mimic interactions in the human body [19]. Generating this realism often involves the employment of advanced microscale fabrication techniques, including 3D printing, compressive buckling, and kirigami-based folding to create the necessary spacing between organoids for their growth and reproducibility, as well as organoid organization and size [20–23]. In this section, we will discuss recent microfluidic technologies that advance precision delivery, sustained perfusion, and integrated multi-organ systems.

## 2.1 Reproducible Formation

The development of organoid-on-a-chip devices has significantly advanced the precision and speed of organoid formation. In particular, polymer-based microfluidic chips offer both high customizability and predictable liquid handling. These systems can be patterned with microscale features using techniques, including photolithography, two-photon lithography, micro-milling, and 3D printing. Computational tools like COMSOL allow for the fine-tuning and prediction of fluid dynamics, along with properties like temperature distribution, concentration gradients, and flow velocity.

Such precisely engineered platforms address key challenges in organoid isolation and reproducibility by enabling the prediction and control of cell aggregation in designated features—such as microwells—over time. This is in contrast to traditional bulk spheroid formation methods that often produce heterogeneous populations, requiring large sample sizes to achieve statistical significance—an inefficiency that poses challenges in clinical and pharmaceutical contexts.

One notable device by Prince et al., aiming to rapidly form large numbers of spheroids with minimal handling, utilized polydimethylsiloxane (PDMS) microfluidic chips with microwells to generate MCF-7 spheroids of uniform diameter (100 or 300 µm) within 72 hours (Fig. 3A, B). The chip design also incorporates a tunable secondary layer of cascading microfluidics that creates a concentration gradient of doxorubicin (DOX) across samples for rapid screening of dosage efficacy on cancer cells [24]. This PDMS microwell system was further applied to patient-derived breast cancer organoids to evaluate the patient-specific efficacy of Eribulin (Fig. 3C) [24]. A similar microwell-based design enables the formation of pancreatic duct-like organoids with a consistent size (~50 µm in diameter) over a 30-day period [25]. The spatial separation of microwells limits interaction between organoids, improving reproducibility and enabling high-quality imaging by forming organoids through shear-force and gravity-induced cell localization.

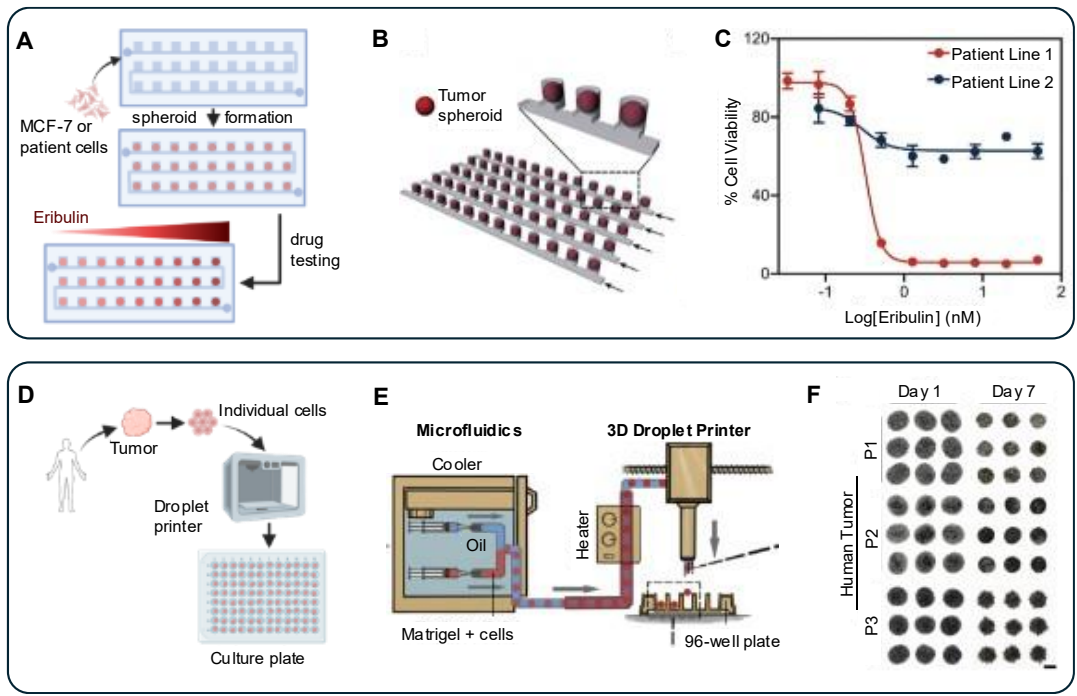

**Figure 3: Reproducibility-targeted microfluidic platforms using microwell arrays and automated organoid generation.** *Microwell systems:* **A)** Experimental workflow for MCF-7 or patient-derived spheroid formation with Eribulin stratification. **B)** PDMS chip layout to generate uniform-sized spheroids within a 72 hour time frame [24]. **C)** Patient-specific tumor spheroid viability in response to Eribulin exposure [24]. *Automated Generation:* **D)** Schematic of biopsy processing to a 96-well culture plate. **E)** Microfluidic droplet printer design [26]. **F)** Growth of patient-derived organoids over 7 days [26]. Parts **A, D** created in BioRender.com. Parts **B, C** used with permission from Wiley-VCH GmbH (2021). Parts **E, F** used with permission from Cell Reports (2020).

Beyond chip-based systems, another promising approach for achieving organoid reproducibility and rapid precision forming is droplet organoid generation [27]. These droplets generally consist of an inner culture chamber with cells that will form the organoid and a shell layer that is typically produced from alginate or collagen. Formation of these droplets can be achieved through microfluidic planar focusing or automated droplet printing methods. Figure 3D shows an example of using a 3D droplet printer to generate Matrigel-embedded patient-derived organoids in a 96-well plate [51]. The 3D printing process leverages thermally controlled phase transitions to form droplets that solidify prior to printing, as shown in Fig. 3E, which creates nearly identical droplets in size and composition. The resulting organoids retained patient-specific characteristics like size, structure, and protein expression that replicate the native tissue (Fig. 3F). Several comprehensive reviews focusing on droplet-based organoid technologies are available [27,28]. While droplet systems have the benefit of creating large quantities of uniformly sized organoids rapidly, limitations include restricted growth time and the difficulty of introducing differentiation factors without removing the protective shell layer.

Both chip-based and droplet-based organoid culture systems offer cost-effective solutions by minimizing reagent consumption during culture [29]. Notably, organoid-on-a-chip platforms also enable continuous nutrient perfusion and waste removal, which supports long-term viability and physiological relevance. This advantage leads into the next discussion: organoid perfusion chips, which use compact, localized microfluidic designs to further enhance culture conditions and enable real-time monitoring through integrated sensors.

## 2.2 Enhancing Perfusion Strategies for Organoid Culture

To address the limitations of nutrient and oxygen delivery in 3D organoid cultures, various perfusion strategies have been developed. Among these, micropillar-based systems, which can either function solely as a mechanism for spacing (typically made of solid and rigid PDMS) or serve as permeable, synthetic vasculature (made of soft, polytheylene glycol diacrylate) for organoids to grow around, have shown promise [30,31]. The inter-pillar spacing plays a critical role in regulating the size of the organoid, with one example supporting the formation of 200 μm spheroids that remain isolated for up to 30 days of culture [32]. Beyond controlling size, the spacing of organoids also simplifies the ease of individual imaging and minimizes the inter-organoid interactions during differentiation, improving nutrition consistency [30,33]. Grebenyuk et al.

demonstrated the use of polyethylene glycol diacrylate (PEGDA) micropillars in a perfused grid-like array for *in-situ* differentiation and maturation of organoids (Figures 4A, 4B). Fluorescence staining for apoptotic factor Caspase 3 and hypoxia marker H1F1-α in Figure 4C showed reduced cell death and oxygen limitation, suggesting optimal nutrient and oxygen access. Similar staining after 62 days under flow conditions, showed enhanced maturation and native brain tissue fidelity in these organoids [31]. Overall, the increased perfusion and spatial arrangement in these pillar systems optimize nutrient and oxygen delivery, waste removal, and size consistency, improving the reproducibility of properties between organoids.

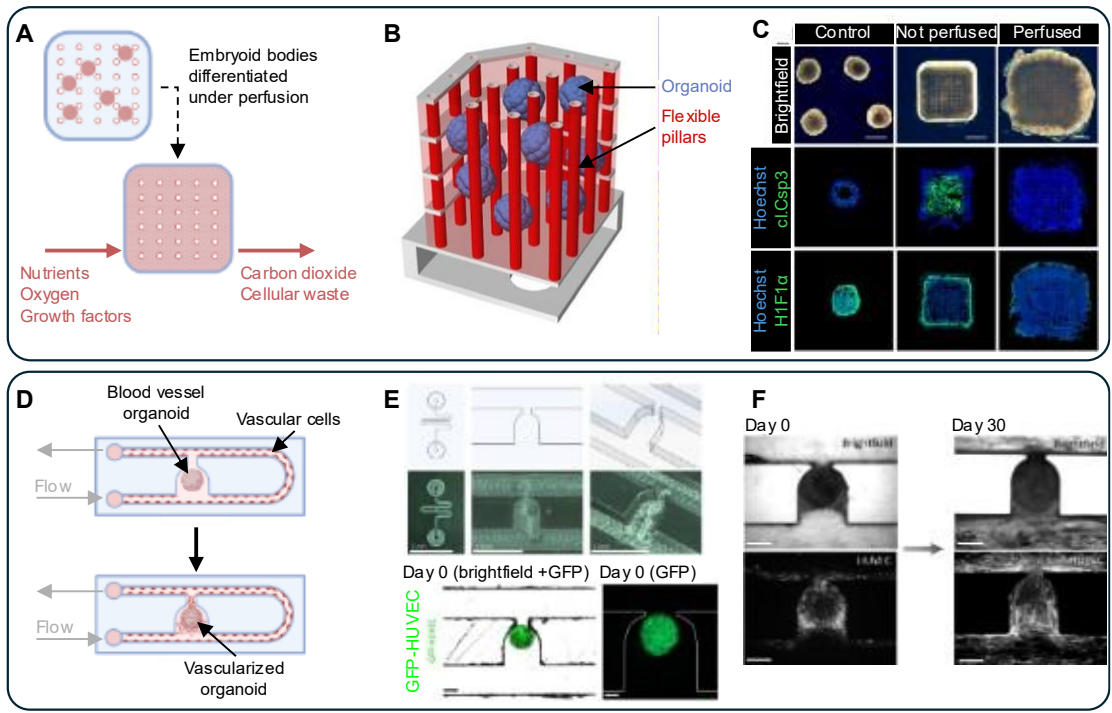

**Figure 4: Approaches to improving perfusion through organoids. Micropillar system: A)** Formation overview from seeded embryoid bodies to fully developed tissue with perfusion. **B)** Flexible, permeable polyethylene glycol diacrylate (PEGDA) micropillars capable of providing synthetic vascularization [31]. **C)** Visualization of tissue growth with and without perfusion through the flexible pillars, showing the lack of cl.Csp3 around the center and limited presence of H1F1α after 62 days [31]. **Vascularization. D)** Overview of the vascular incorporation process described by Quintard et al. **E)** Blood vessel organoid (BVO) positioned in a cyclic olefin copolymer (COC) chip prior to incorporation of human umbilical vein endothelial (HUVEC) cells [34]. **F)** Imaging of fully vascularized blood vessel organoids (green) with tracking of beads (red) through the structures [34]. Parts **A, D** created in BioRender.com. Parts **B, C** used with permission from Nature (2023). Parts **E, F** used with permission from Nature (2024).

An alternative approach to designing perfusion systems is the co-culture of organoids with vascular cells, which establish natural perfusion pathways both on the surface and within the organoid. By creating branching networks, these vascular cells provide biologically relevant perfusion structures for nutrient transport, addressing the limitations of nutrient access within the organoid core [35]. Improved nutrient delivery supports the growth of larger organoids and enables longer culture durations, ultimately promoting the development of mature tissue features and enhancing disease modeling. For instance, vascularized pancreatic organoids exhibit luminal structure formation, fluid secretion, and disease traits consistent with cystic fibrosis [36]. Vascular cells can initially be separated from the organoids using a semi-permeable membrane in a multi-chamber model while being connected to the nutrient inlet supply [37,38]. Alternatively, vascular cells can be co-cultured directly with organoids in the same culture chamber, as demonstrated in blood vessel organoids (Figure 4D) [34]. In this design, a goblet-like chamber houses the organoid (Figure 4E), while human umbilical vein endothelial cells (HUVECs) are flowed through adjacent microchannels, subsequently attaching to the chamber surfaces and forming vascular networks. Flow of nutrients throughout the organoid was demonstrated by tracking fluorescent beads of 1 µm diameter—much larger than most proteins or sugars—circulating throughout the structure (Figure 4F). Despite its effectiveness in delivering nutrients to the organoid core, this vascular co-culture strategy presents challenges for subsequent analyses, such as isolating the target organoid for optical evaluation. Moreover, co-culture vascularization is better suited to the development of spheroids that are already differentiated, where precise spatial control of growth factors is less critical.

## 2.3 Systemic Interactions

Co-culture microfluidic chips can also be designed with multiple chambers separated by a permeable membrane, allowing for cross-talk between different types of organoids to study systemic responses. Figures 5A and 5B show an example of such a device, which integrates microwells with separated chambers to examine the unintended side effects of the antidepressant clomipramine on cardiac organoids [39]. In this setup, the chambers are separated by a polycarbonate permeable membrane: liver organoids metabolize the drug, and the resulting byproducts are introduced to the cardiac organoids. As shown in Figure 5C, cardiac organoids in the co-cultured system including liver organoids exhibited more significant damage, highlighting the importance of multi-organ drug testing to determine effective and safe treatment concentrations for patients.

On-chip designs of microfluidic exchange, are crucial for modeling multi-organ interaction models and hold promise for ultimately reducing or replacing animal testing.

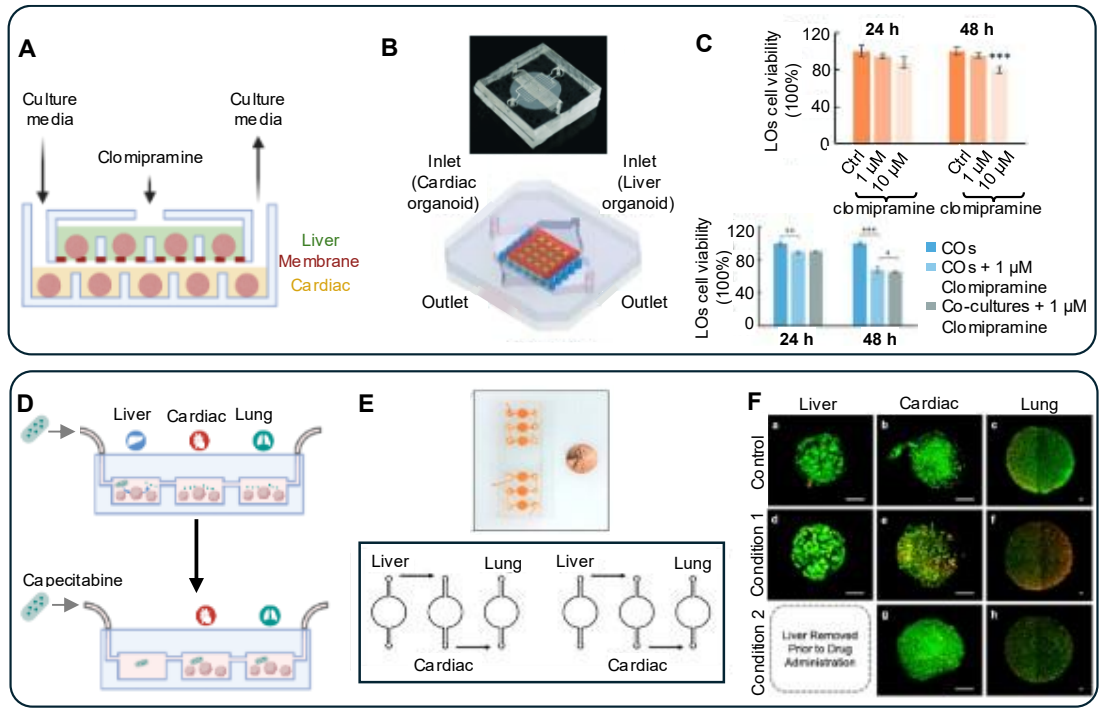

**Figure 5: Systemic interaction models. Multi-chamber system: A)** Side-view schematic showing permeable membrane-separated culture chambers for liver and cardiac organoids to study multi-organ effects of clomipramine. **B)** Images of the overall device layout and assembled PDMS device with polycarbonate porous membrane separation [39]. **C)** Viability of liver organoids (LOs) and cardiac organoids (COs) after 24 and 48 hours of clomipramine treatment, showing significant cardiac effects. **Body-on-a-chip system: D)** Schematic overview of a three-organ chip with capecitabine treatment to show the necessity of multiple organoids in a system. **E)** Image of the three-organoid adhesive film-based (AFB) microfluidic chip connected through external tubing [40]. **F)** Imaging results showing damage to cardiac and lung tissue in a system containing liver organoids, which does not occur in their absence [40]. Parts **A, D** created in BioRender.com.[39] Parts **B, C** used with permission from The Royal Society of Chemistry (2021). Parts **E, F** used with permission from Elsevier (2020).

Multi-well microfluidic models can also be used to study the broader impacts of cross-organ toxicity and incorporate more than three organoid types for body-on-a-chip studies. These platforms enable nutrient and byproduct exchange between different organs, representing an essential step toward comprehensive in vitro drug evaluations as alternatives to animal studies. For instance, drugs primarily targeting lungs can have unexpected adverse effects on other organs, most commonly the heart, brain, and liver [40,41]. A flexible adhesive film-based chip design shown in Figure 5D demonstrates this principle. The chip was first patterned by a computer-controlled razor plotter and then folded to create channels with a height of 560 µm from four stacked layers. The film structure is sandwiched between a glass slide base and a laser-cut polymethyl methacrylate (PMMA) cover, with external tubing connecting the chambers (Figure 5E). Using this system, the cross-toxicity of the drug capecitabine was assessed by monitoring liver, heart, and lung organoids. As shown in Figure 5F, the metabolic breakdown of capecitabine by the liver organoids increases toxicity in heart and lung organoids compared to systems without a liver organoid [40]. The platform was also expanded to include six organoid types—liver, heart, lungs, blood vessels, testes, and brain—to evaluate ifosfamide toxicity. Similar outcomes were observed, with systems containing liver organoids exhibiting increased toxicity, suggesting the necessity of metabolizing tissues in multi-organ drug assessments. Importantly, as the number of organoids increases, tailored culture media and growth protocols are often required to sustain the entire system. The complexity of body-on-a-chip designs is crucial for maintaining both safety and accuracy in drug testing, offering a pathway to reduce reliance on animal models.

Overall, microfluidic technologies enhance organoid models by enabling reproducibility, deeper perfusion, and more physiologically relevant interactions. Leveraging these designs also allows integration of diverse monitoring modalities, including imaging, genetic profiling, and sensor-based readouts.

## 3. Organoid Sensor Interfaces

In parallel with highly tunable microfluidic systems, a wide range of sensor technologies have been developed to track organoid development and function. Sensors integrated into developmental or drug-response models must exhibit minimal interference with organoid growth, maintain stable performance under culture conditions, and provide sufficient spatial resolution to capture structural and functional complexity. Ideal designs include non-destructive, temporally resolved monitoring of the same organoid over extended culture periods—often several months—without restricting organoid development [15]. While multiple sensing targets have been explored, most studies focus on either physical properties or chemical markers related to metabolism and signaling [42,43]. Incorporating features such as microscale dimensions or mechanical flexibility allows sensors to interface with organoids while minimizing structural damage and developmental disruption.

## 3.1 Physical Sensors

Physical sensors measure characteristics of the culture environment or the organoid itself, such as temperature, pressure, electrical signals, or displacement caused by organoid growth. These sensors can be positioned on the culture vessel [44,45], embedded inside the organoid [46–48], or wrapped around the organoid surface [49–52]. They are particularly suited for organoid types with intrinsic electrophysiological activity, such as brain [53–55] and heart organoids [44,56,57]. Because electrophysiological properties can be affected by pharmaceutical agents, drug response is a common validation method for such sensors, where firing rate or beating rate may have a proportional response to drug concentration.

The Rogers group developed a flexible multielectrode array (Figure 6A) that enables spatial monitoring of electrophysiological signals in co-cultured cortical and astrocyte spheroids. This device uses 25 low-impedance electrodes to map the field potential across the organoid surface [49]. Its adaptable and basket-like shape is formed by compressive buckling of photolithographically patterned 2D gold electrodes on a pre-strained PDMS substrate as an assembly platform (Figure 6B). Using this basket, spatial mapping of electrophysiological activity revealed firing rate reductions following administration of tetrodotoxin (TTX), a sodium channel blocker (Figure 6C). These results underscore the importance of developing monitoring methods capable of capturing spatial heterogeneity across the spheroid structure, rather than relying on single-point or planar electrodes. In the context of assembloids, such methods are particularly valuable for revealing electrical connections across distinct brain regions.

A wrapping array of a similar concept was designed using a kirigami-inspired buckling approach, creating a thinner suspended basket shape to monitor either single organoids or assembloids as a part of the growth vessel (Figure 6D). This array employs a spiral design of platinum and gold patterned between two layers of SU-8 followed by etching of a sacrificial layer on the silicon wafer to release the basket structure (Figure 6E). The device contains 32 impedance electrodes that can simultaneously record local signals from suspended neural organoids at a sampling rate of 20 kHz, enabling stable long-term monitoring over 179 days of culture [50]. The high density of contact points enhances spatial resolution, while the flexible design provides a high level of temporal resolution. Drug validation confirmed the increase in the firing rates of the cortical organoids after administration of 4-AP, consistent with expected pharmacological effects (Figure 6F). The system also demonstrated responsiveness to optogenetic stimulation. Importantly, its thin, net-like architecture allowed optical imaging of suspended organoids without interference, making it compatible with established evaluation methods.

Together, these advances demonstrate that thin, flexible electronic suspension systems can integrate seamlessly with organoid growth and function. By offering minimally invasive, high-resolution, and long-term monitoring, they represent a significant step toward more physiologically relevant readouts in organoid research.

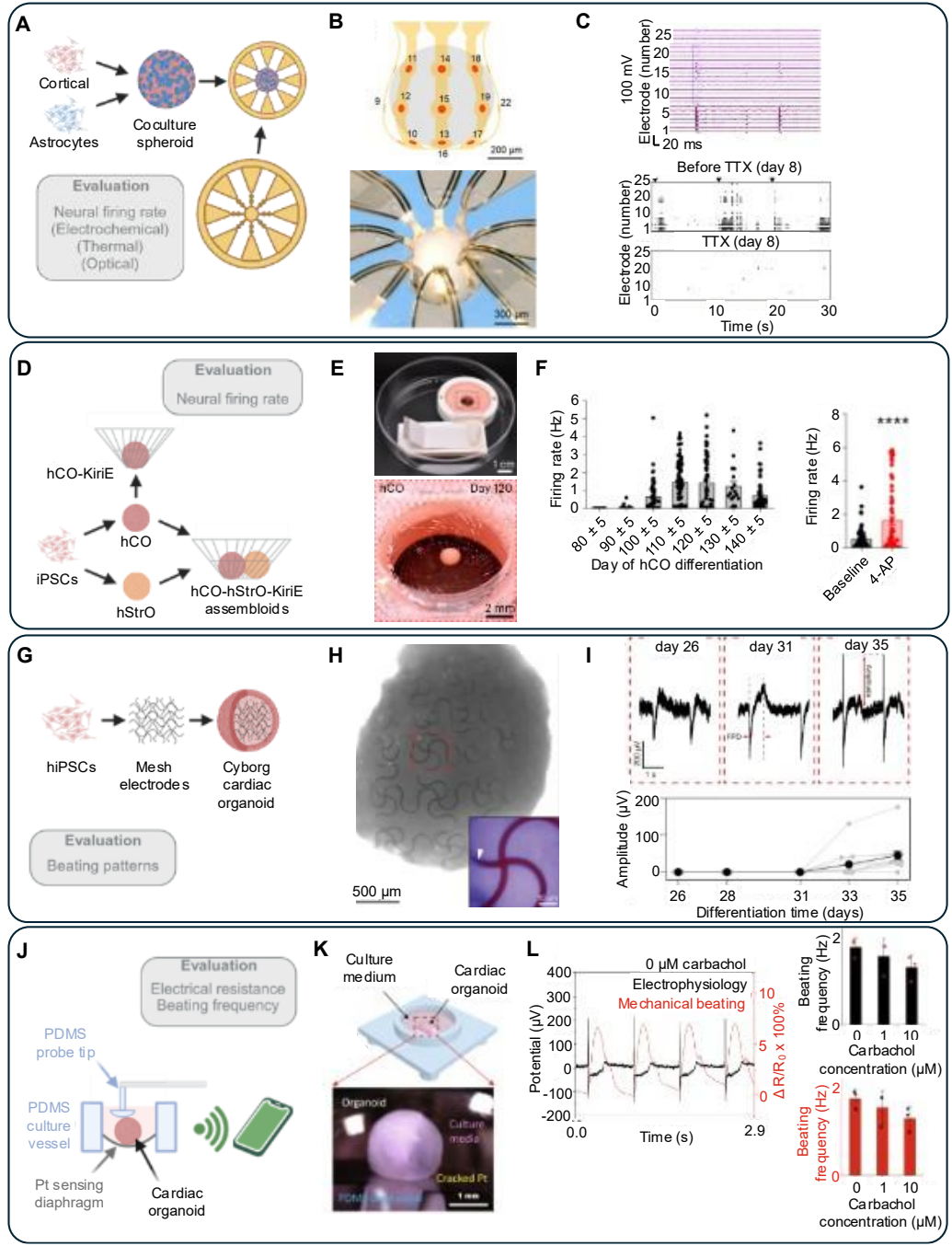

**Figure 6: Physical sensing methods for organoid analysis. Multifunctional wrap: A)** Multifunctional mesoscale framework (MMF) for spatial recording of electrophysiological activity on cortical and astrocyte co-culture neural spheroids, integrating microelectrodes with thermal, optical, and electrochemical sensors. **B)** Design of the electrode array and holding structure around the pre-formed spheroid [49]. **C)** Multichannel electrophysiological firing recordings around the spheroid before and after exposure to tetrodotoxin (TTX) [49]. **Wrapping basket: D)** Experimental overview of differentiated cortical organoids (hCO) and striatal organoids (hStrO) analyzed individually or as assembloids via electrodes. **E)** Kirigami-inspired 3D electronics structure design with microelectrodes to collect electrophysiological measurements on neural organoids [50]. **F)** Comparison of firing rates during hCO

differentiation, as well as at 138 day before and after exposure to 4-AP, a known stimulant [50]. **Internal mesh: G)** Process of cardiac organoid formation around a flexible mesh structure *in vitro*. **H)** Image of mesh embedded within a cardiac organoid after 40 days of differentiation [48]. **I)** Beating amplitude over time during cardiac organoid differentiation [48]. **Diaphragm stretching: J)** Side view demonstration of a PDMS holder with tip and Pt sensing diaphragm for remote monitoring of cardiac beating activity. **K)** Device image of an organoid on a cracked Pt diaphragm [57]. **L)** Electrophysiological (black) and mechanical (red) recordings of beating activity with increasing carbachol concentration [57]. Parts **A, D, G, and K** were created in BioRender.com. Parts **B, C** used with permission from Science Advances (2021). Parts **E, F** were used with permission from Nature (2024). Parts **H, I** used with permission from American Chemical Society (2019). Parts **K, L** used with permission from Nature (2022).

Beyond external electrode systems, several internal physical sensor designs have been developed for integration within spheroid structures. Examples include internalized RF antennas for remote signal transmission [53,58] and pointed electrode arrays that partially pierce the spheroid to attach it to external measurement devices [59,60]. One notable approach employed a flexible microelectrode array sheet embedded within cardiac organoids during growth (Figure 6G and 6H) to monitor beating throughout different developmental stages[48]. These electrodes were fabricated by depositing platinum and poly(3,4-ethylenedioxythiophene) (PEDOT) on a stretched substrate, forming a serpentine structure that buckles upon release of the pre-strain in the substrate [48]. The mesh was connected to an external system for power supply and data acquisition to record signals. This design allowed the continuous growth of the formed "cyborg" organoids over 35 days while capturing electrophysiological signal amplitude and frequency as the cardiac organoid matured (Figure 6I). Similar mesh electrodes from this group were adapted with graphene barcoding to spatially corroborated RNA sequence [47], further enabling a deeper understanding of the formation, characterization, and integration of these mesh electrodes [46]. Like suspension meshes, these systems maintain optical transparency, allowing imaging through the organoid structure. The successful incorporation of spatially distributed internal electrodes highlights the potential for minimally disruptive monitoring of electrophysiological and optical properties in growing organoids.

While electrical properties are most commonly measured in the literature, the development and function of some organoids also rely on mechanical forces [42,61]. The mechanical properties of the surrounding environment can be evaluated using bulk rheology [62]. they are currently quantified in individual organoids using techniques like optimal tweezers, parallel plate potentiometry, laser ablation, or atomic force microscopy (AFM) [63,64]. An integrated approach combining mechanical and electronic sensing is demonstrated in Figure 6K, where a PDMS tip coupled with a platinum-coated sensing diaphragm is used to track cardiac organoid beating. The culture vessel shown in Figure 6L simultaneously collected both electrophysiological signals and mechanical beating data, demonstrating that treatment with carbachol caused a reduction in beating frequency, which can be monitored remotely through both modalities (Figure 6M). Similarly, a microneedle array coupled with a deformable diaphragm has been used to the ability to simultaneously track both regular and interrupted beating patterns using a real-time monitoring system [65]. Because general movement may interfere with mechanical signals, complex mapping and convolution are often required to isolate true mechanical measurements and provide adequate spatial resolution. Nevertheless, electro-mechanical coupling systems already have broad applications in drug screening. These systems not only indicate when beating becomes prominent in cardiac organoid development, but also reveal alterations to heart rhythm after exposure to common chemotherapeutic treatments, closely recapitulating arrhythmias observed in patients.

## 3.2 Chemical Sensors

Chemical sensors are used to monitor molecular changes either in the surrounding environment or within the organoids themselves, tracking parameters such as pH, oxygen concentration, and biomarker levels. Monitoring these factors offers analytical insights into the culture process for both research and industrial applications. While in industrial settings, culture monitoring is typically performed by optical sensors or spectroscopic analysis [66], research applications often require more detailed analysis, given the high sensitivity of differentiated organoid models to growth factors and environmental conditions.

Many on-chip microfluidic systems can employ in-line chemical monitoring, where a portion of the culture media is isolated for analysis either in real time or through periodic sampling. In-line chemical sensors often focus on detecting biomarkers that can be quantified using existing cellular assays, such as oxygen, pH, or specific secretions linked to the organ of interest [67–69]. For example, hepatic organoids are monitored for transferrin and albumin, both key proteins synthesized by mature liver tissue [41,70], while neural organoid maturity can be assessed through neurotransmitters like glutamate or dopamine [55,71]. Another example of an organ-specific biosensor target is creatine kinase (CK-MB), an isoenzyme linked to cardiac cell damage, which can be monitored using an aptamer-based sensor (Figure 7A). In this design, media from the culture chamber was diverted to the sensor through separate microfluidic connections (Figure 7B), enabling measurement of CK-MB concentration variations in response to doxorubicin dosage and exposure time (Figure 7C) [72]. These in-line sensors can reduce the need to terminate cultures for endpoint analysis, such as fluorescent assays, immunohistological imaging, or flow cytometry [73]. However, in-line sensing is a representation of the entire organoid environment, rather than capturing spatial variations around an individual organoid, and additionally, biorecognition elements like aptamers often have limited stability over time. Despite these limitations, combining in-line chemical monitoring with other physical readouts, such as temperature, provides a more comprehensive view of growth conditions, enabling precise control and understanding of organoid development.

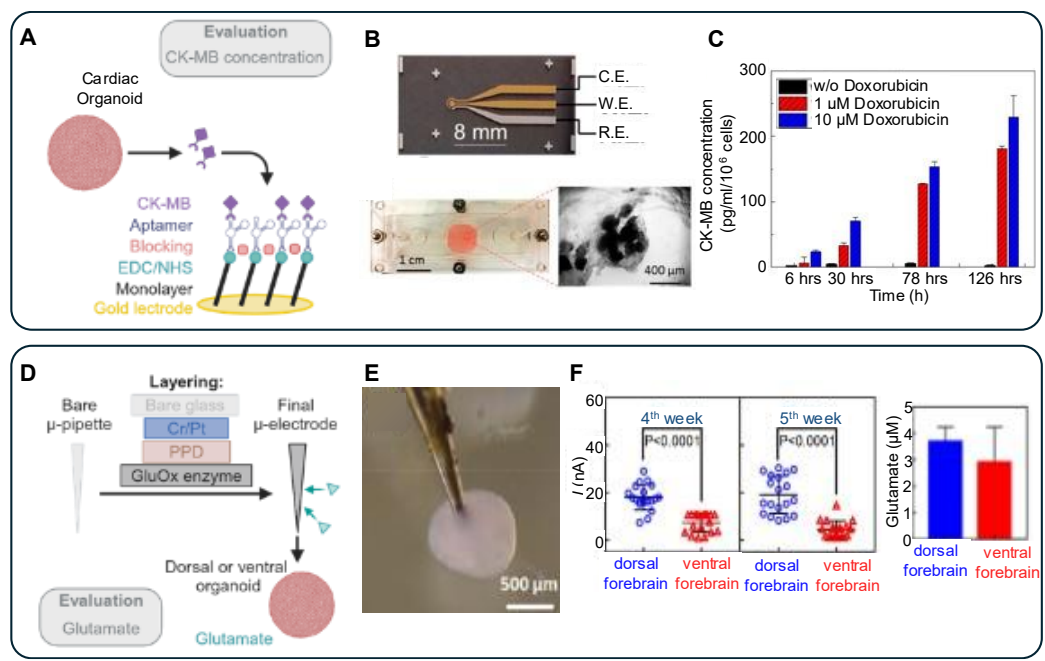

**Figure 7: Chemical sensing strategies for organoids. Bulk solution sensing: A)** Aptamer-based electrochemical sensor for detecting CK-MB in damaged cardiac organoids. **B)** Integration of the sensor with organoid culture chamber on a PDMS chip [72]. **C)** Measurement of cardiac organoid damage with increasing exposure of doxorubicin [72]. **Penetrating electrode: D)** Functionalization process of a penetrating electrode for sensing glutamate in forebrain organoids. **E)** Image of the electrode inserted into neural organoid tissue [55]. **F)** Differences in electrical current and associated glutamate levels between dorsal and ventral forebrain organoids at different lengths of culture time [55]. Parts **A, D** created in BioRender.com. Parts **B, C** used with permission from American Chemical Society (2016). Parts **E, F** were used with permission from MDPI (2018).

Sensing chemical concentrations within singular organoids, rather than in the bulk solution, requires more localized spatial electrode designs like permanently implanted sensors [74] or temporary microelectrode punctures [55]. For example, Nasr et al. employed a temporary penetrating electrode to monitor glutamate levels in neural organoids differentiated to model different brain regions. As shown in Figure 7D, a glass pulled tip was coated with Cr/Pt, followed by drop casting of the enzyme glutamate oxidase (GluOx) to selectively detect glutamate. The electrode tip was inserted into brain organoids periodically using a micromanipulator controller, enabling electrical monitoring of regional glutamate dynamics [55]. Measurements over the course of differentiation revealed higher levels of glutamate in dorsal forebrain organoids compared to ventral forebrain organoids, consistent with *in vivo* brain architecture (Figure 7F) [55]. Tracking the biochemical properties of individual organoids over time can generate data-driven insights into the development of their size, physical properties, and chemical information.

Non-destructive, real-time monitoring of chemical signals from the same organoid over time is key to reproducible measurements, system standardization, and validation of biological models using sensors. Real-time measurements have been demonstrated in breast cancer stem cells differentiating into spheroids, where sensors monitor the lactate and oxygen metabolism in real-time throughout the growth process [75]. As chemical sensing techniques continue to advance, they offer increasing robust non-destructive, real-time monitoring, making them particularly suited to the specific needs of organoid research and applications.

## Recent Organoid Applications

Organoid models are highly valuable for their ability to recapitulate physiologically relevant structures and responses, especially in testing pharmaceutical agents that can facilitate cellular growth or model diseases. The 3D organization of organoids more accurately reflects *in vivo* behavior compared to traditional cell layer models, due to enhanced cell-to-cell interactions and the presence of extracellular matrix. Here, we highlight representative recent applications that leverage advanced microfluidics or microelectronics systems. As illustrated in Figure 8, these applications generally fall into three main categories: drug delivery, personalized medicine, and developmental studies.

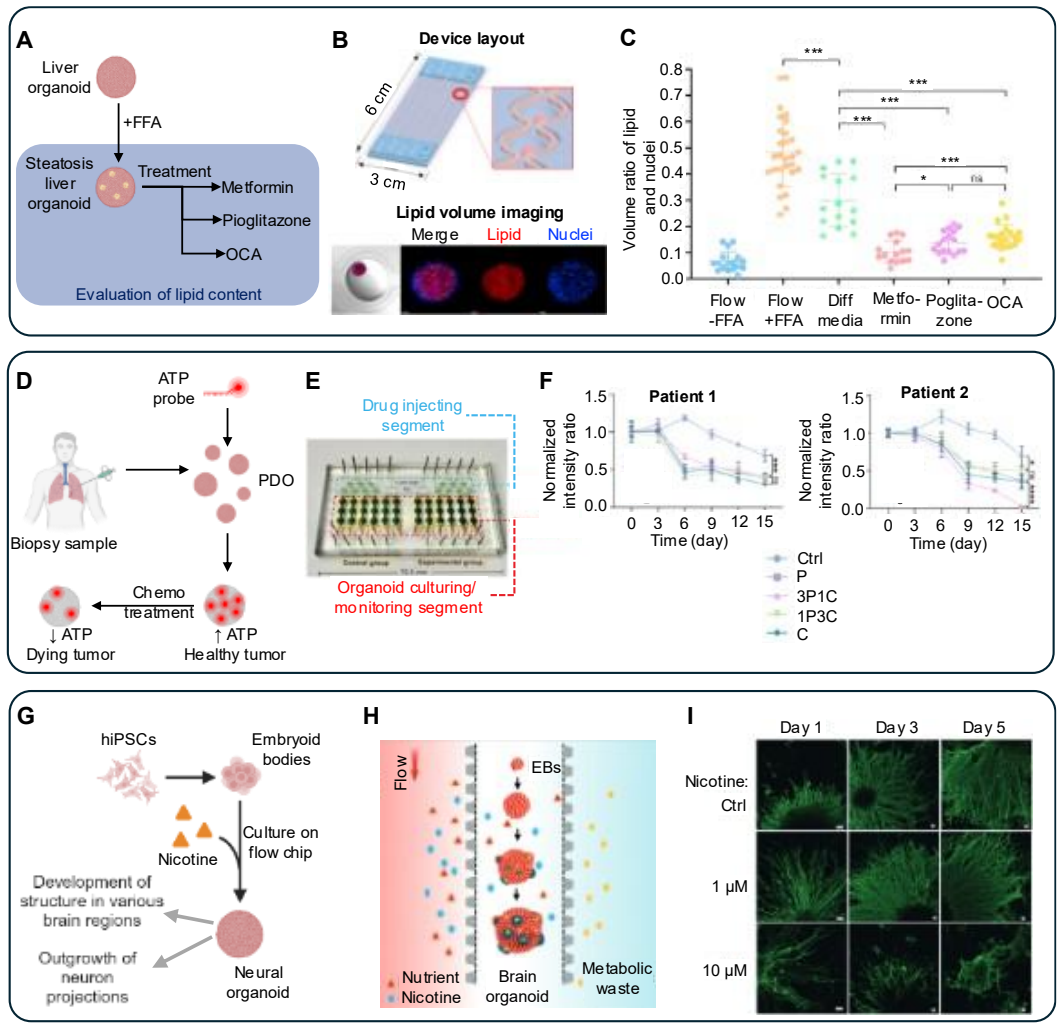

**Figure 8: Notable recent application examples in drug delivery, personalized medicine, and developmental biology. Drug delivery. A)** Schematic demonstrating induction of steatosis in liver organoids cultured in SteatoChip device using flufenamic acid (FFA) and subsequent drug testing. **B)** Device layout to generate native flow patterns to liver organoids and example fluorescent stain to compare overall volume properties with proportional lipid quantity [76]. **C)** Evaluation of Metformin, Pioglitazone, and OCA effect on reducing volume of lipid when compared to the Flow+FFA *in vitro* steatosis effect organoids [76]. **Personalized medicine. D)** Experimental flow of generation of patient-derived organoids (PDO) and utilization of ATP optical sensing for testing of chemotherapeutic treatments for tumors. **E)** Device layout for optimized culture of multiple PDO wells and optical sensor design using activated fluorescent Cy3 upon ATP binding [77]. **F)** Comparison of chemotherapeutic compound combinations for two patients, showing different optimal treatments [77]. **Developmental study. G)** Summary of study for applying nicotine to developing neural organoids in a flow system to simulate nicotine consumption during pregnancy. **H)** Device layout showing directional flow of nutrients, nicotine, and removal of metabolic waste products [78]. **I)** Fluorescent imaging of neuronal outgrowth from formed organoids over 5 days of exposure at different concentrations of nicotine [78]. Parts **A, D, and G** created in BioRender.com. Parts **B, C** used with permission from Elsevier (2021). Parts **E, F** used with

permission from American Chemistry Society (2024). Parts **H, I** used with permission from The Royal Society of Chemistry (2018).

## 4.1 Drug Screening

Effective drug screening relies on identifying appropriate dosages and exposure times to safely move toward clinical trials. In the pharmaceutical industry, there remains a critical need for predictive models to reduce the expenditure of resources on drugs that perform well *in vitro* or animal studies but fail in human clinical trials - a failure rate as high as 90%, with costs of $1-2 billion per drug brought to market [79]. Compared to animal models, human organoids offer an ethical and more predictive platform for testing drug efficacy in human patients. For example, Teng et al. developed a microwell device known as SteatoChip, designed to simulate steatosis, a disease that causes excessive fat accumulation in the liver [76]. The condition of steatosis was reproduced in hepatic organoids by adding flufenamic acid (FFA), followed by the examination of the efficacy of several drugs (Figure 8A). By employing fluorescent imaging to compare the lipid volume with the overall volume of the organoid (Figure 8B), the performance of metformin, pioglitazone, and OCA in reducing lipid concentration was demonstrated (Figure 8C). The study found that metformin most effectively reduced steatosis in liver organoids, consistent with its known therapeutic effectiveness in the human body. The high-throughput chip format not only simulated perfusion conditions in the liver but also enabled the simultaneous culture of large numbers of spheroids, providing a rapid and robust platform for drug testing.

Additional studies comparing 2D and 3D models of various organs, including the breast [80], colon [81], and liver [82], have shown that 3D models yield more accurate predictions of drug responses and patient survival. Among available platforms, microwell arrays are particularly effective for drug screening, as they enable the controlled generation of large numbers of spheroids and allow for precise exposure to drug concentrations. Drug evaluation typically involves assessing organoid size, viability, and specificity to the target organ, and often incorporates targeted assays for markers of cell death or metabolic activity. In this context, electrochemical sensors have emerged as powerful tools for real-time monitoring. Rapid formation and timely evaluation of organoids are essential to identifying effective treatment within clinically relevant timeframes, both for general drug screening and for personalized medicine applications.

## 4.2 Personalized Medicine

Capturing the drug response of organoids is highly valuable not only for evaluating a drug's therapeutic potential but also for predicting patient-specific outcomes. Patient-derived organoids represent a significant step toward personalized medicine, especially for optimizing the efficacy of chemotherapy treatments for cancer patients [83,84]. Patient-derived organoid models have the benefit of reproducing tumor properties like protein expression or structural variations, enabling a detailed examination and simulation of patient responses to chemotherapeutics or combination therapies. These models have proven effectiveness in modeling cancers of the liver [85,86], pancreas [87–89], breast [90], and glioblastoma [91], among others. However, there remains a critical need to develop faster and more reproducible culture systems to shorten the time from patient biopsy to robust analysis for clinical decision-making and, ultimately, to the application of the therapeutic agent back to the individual patient.

A recent study by Zhang et al. demonstrated the feasibility of rapidly generated, personalized organoids to monitor patient-specific metabolic changes due to combination chemotherapy using an optical ATP sensor (Figure 8D). The sensor incorporated Cy3, which fluoresces upon binding to ATP-targeted aptamer strands, producing an intensity signal proportional to ATP presence. The sensor molecules were injected into the microfluidic device that enabled both culturing and imaging organoids in the same integrated platform (Figure 8E). The results showed a reduction in metabolism across all treatments, with the combination 1P3C being more effective for patient 1, while 3P1C was more effective for patient 2 (Figure 8F). This study highlights the potential of personalized treatment modelling within 15 days of biopsy sampling, a timeframe that aligns with clinical intervention needs, by combining microfluidics with optical sensing for streamlined monitoring.

## 4.3 Developmental Studies

Fundamental questions remain about organ development during fetal growth and the cellular processes underlying tissue regeneration and remodeling in the adult body. *In vitro* biological models have been used to identify remodeling processes, like the hematopoietic function of bone marrow [92] and immunological relationships, like the role of macrophages in the development of fetal brain tissue [93]. Organoid models derived from differentiated stem cells offer a window into organogenesis, enabling the study of the formation of substructures that are critical to organoid function. Notable examples include the formation of wrinkling patterns in brain organoids [94] and the formation of islet cell groups in pancreatic organoids [95]. As illustrated in Figure 8G, a microfluidic platform was developed to investigate the impact of nicotine on neural organoid development over five days. In this flow chip system, nicotine and nutrients were perfused into preformed embryoid bodies (EBs), while metabolic waste was simultaneously removed (Figure 8H). Subsequent fluorescent imaging revealed reduced neural outgrowths from nicotine-exposed organoids, indicating that even short-term exposure to high concentrations of nicotine is highly disruptive to neural development (Figure 8I). Additional imaging further identified structural alterations within the brain organoids. The in vitro visualization of the developmental disruption provides a tangible and ethical approach to studying the effects of nicotine exposure during gestation.

A key consideration in designing organ development models is achieving long-term culture and sufficient perfusion to deeper layers of the organoid to support maturation. Longevity is particularly important, as human organ development spans at least nine months, and tissue remodeling occurs throughout the human lifespan. Generating suitable microfluidics to be incorporated within organoids remains challenging, especially in creating appropriately sized structures for synthetic vascularization to ensure effective perfusion. Organoid development is typically assessed by evaluating the identity, structure, and function of the organoid, often using immunohistology of relevant proteins. However, these methods are disruptive, requiring fixation or sectioning of organoids for imaging. Sensor-based alternatives that target biomarkers offer a promising noninvasive approach to monitor developmental markers. Current organoid models are currently limited to a few months of growth, halting tissue maturation at an early stage. Future research will need to focus on developing long-term, noninvasive culture and monitoring methods to more effectively recapitulate organ development.

## 5      Future Considerations

While significant steps have enhanced temporal and spatial control in organoid engineering, current technologies still cannot fully replicate the phenomena observed in human tissues. A major limitation is the inability to grow and evaluate organoids to levels of maturity and complexity that accurately reflect human biology. Brain development, for example, unfolds over decades, and much remains to be discovered about ways to quantify this progression in terms of both structural architecture and neural connectivity [96]. Model longevity is particularly critical when modeling neurodegenerative diseases such as Alzheimer's, dementia, or general neurodegeneration – conditions that manifest after decades of brain development and remodeling. Longevity has been approached biologically utilizing CRISPR for genetic reprogramming [97] and through model regeneration through organoid passaging and expansion [98]. Achieving appropriate culture longevity will rely on creating systems with stable long-term perfusion and sensing capabilities using the careful selection of model, fabrication methods, and material type, to tackle the challenge of the size scale required for appropriate spatial resolution. Combining these biological model enhancements with microfluidic and microelectronic systems may further empower organoids to reach their full potential to reach complexity.

Interdisciplinary approaches that leverage materials science, micromanufacturing, and biomedical engineering will be crucial in advancing healthcare understanding and treatment options using organoids. Together, these efforts promise to advance our understanding of human development and disease, improve predictive drug screening, and accelerate the translation of organoid-based platforms into personalized medicine and therapeutic applications.

**Notes**
The authors declare no conflict of interest.

**Funding source**
Research reported in this publication was partially supported by the National Institutes of Health (USA) under Award Number R21EB033495 (to XW, YZ, and YL), R01MH128721 (to YZ), and R01NS125016 (to YL). The content is solely the responsibility of the authors and does not necessarily represent the official views of the National Institutes of Health.

**References**

[1] F. Pampaloni, E.G. Reynaud, and E.H.K. Stelzer, "The third dimension bridges the gap between cell culture and live tissue," Nat Rev Mol Cell Biol **8**(10), 839–845 (2007).

[2] M. Vinci, S. Gowan, F. Boxall, L. Patterson, M. Zimmermann, W. Court, C. Lomas, M. Mendiola, D. Hardisson, and S.A. Eccles, "Advances in establishment and analysis of three-dimensional tumor spheroid-based functional assays for target validation and drug evaluation," BMC Biol **10**, 29 (2012).

[3] V. Velasco, S.A. Shariati, and R. Esfandyarpour, "Microtechnology-based methods for organoid models," Microsyst Nanoeng **6**(1), 1–13 (2020).

[4] P.-J.H. Zushin, S. Mukherjee, and J.C. Wu, "FDA Modernization Act 2.0: transitioning beyond animal models with human cells, organoids, and AI/ML-based approaches," J Clin Invest **133**(21), e175824 (2023).

[5] T. Takebe, B. Zhang, and M. Radisic, "Synergistic Engineering: Organoids Meet Organs-on-a-Chip," Cell Stem Cell **21**(3), 297–300 (2017).

[6] G.M. Crane, E. Jeffery, and S.J. Morrison, "Adult haematopoietic stem cell niches," Nat Rev Immunol **17**(9), 573–590 (2017).

[7] X. Yin, B.E. Mead, H. Safaee, R. Langer, J.M. Karp, and O. Levy, "Engineering Stem Cell Organoids," Cell Stem Cell **18**(1), 25–38 (2016).

[8] R.W. Orkin, P. Gehron, E.B. McGoodwin, G.R. Martin, T. Valentine, and R. Swarm, "A murine tumor producing a matrix of basement membrane.," Journal of Experimental Medicine **145**(1), 204–220 (1977).

[9] N. Gjorevski, N. Sachs, A. Manfrin, S. Giger, M.E. Bragina, P. Ordóñez-Morán, H. Clevers, and M.P. Lutolf, "Designer matrices for intestinal stem cell and organoid culture," Nature **539**(7630), 560–564 (2016).

[10] G. Quadrato, T. Nguyen, E.Z. Macosko, J.L. Sherwood, S. Min Yang, D.R. Berger, N. Maria, J. Scholvin, M. Goldman, J.P. Kinney, E.S. Boyden, J.W. Lichtman, Z.M. Williams, S.A. McCarroll, and P. Arlotta, "Cell diversity and network dynamics in photosensitive human brain organoids," Nature **545**(7652), 48–53 (2017).

[11] S.P. Paşca, P. Arlotta, H.S. Bateup, J.G. Camp, S. Cappello, F.H. Gage, J.A. Knoblich, A.R. Kriegstein, M.A. Lancaster, G.-L. Ming, A.R. Muotri, I.-H. Park, O. Reiner, H. Song, L. Studer, S. Temple, G. Testa, B. Treutlein, and F.M. Vaccarino, "A nomenclature consensus for nervous system organoids and assembloids," Nature **609**(7929), 907–910 (2022).

[12] M. Simian, and M.J. Bissell, "Organoids: A historical perspective of thinking in three dimensions," J Cell Biol **216**(1), 31–40 (2017).

[13] Z. Zhao, X. Chen, A.M. Dowbaj, A. Sljukic, K. Bratlie, L. Lin, E.L.S. Fong, G.M. Balachander, Z. Chen, A. Soragni, M. Huch, Y.A. Zeng, Q. Wang, and H. Yu, "Organoids," Nat Rev Methods Primers **2**(1), 94 (2022).

[14] Y. Gu, W. Zhang, X. Wu, Y. Zhang, K. Xu, and J. Su, "Organoid assessment technologies," Clin Transl Med **13**(12), e1499 (2023).

[15] L. Zhang, L. Wang, S. Yang, K. He, D. Bao, and M. Xu, "Quantifying the drug response of patient-derived organoid clusters by aggregated morphological indicators with multi-parameters based on optical coherence tomography," Biomed Opt Express **14**(4), 1703–1717 (2023).

[16] A. Junaid, A. Mashaghi, T. Hankemeier, and P. Vulto, "An end-user perspective on Organ-on-a-Chip: Assays and usability aspects," Current Opinion in Biomedical Engineering **1**, 15–22 (2017).

[17] S. Velasco, A.J. Kedaigle, S.K. Simmons, A. Nash, M. Rocha, G. Quadrato, B. Paulsen, L. Nguyen, X. Adiconis, A. Regev, J.Z. Levin, and P. Arlotta, "Individual brain organoids reproducibly form cell diversity of the human cerebral cortex," Nature **570**(7762), 523–527 (2019).

[18] J. Drost, and H. Clevers, "Organoids in cancer research," Nat Rev Cancer **18**(7), 407–418 (2018).

[19] H.-Y. Tan, H. Cho, and L.P. Lee, "Human mini-brain models," Nat Biomed Eng **5**(1), 11–25 (2021).

[20] Z. Chen, N. Anandakrishnan, Y. Xu, and R. Zhao, "Compressive Buckling Fabrication of 3D Cell-Laden Microstructures," Advanced Science **8**(17), 2101027 (2021).

[21] H. Luan, Q. Zhang, T.-L. Liu, X. Wang, S. Zhao, H. Wang, S. Yao, Y. Xue, J.W. Kwak, W. Bai, Y. Xu, M. Han, K. Li, Z. Li, X. Ni, J. Ye, D. Choi, Q. Yang, J.-H. Kim, S. Li, S. Chen, C. Wu, D. Lu, J.-K. Chang, Z. Xie, Y. Huang, and J.A. Rogers, "Complex 3D microfluidic architectures formed by mechanically guided compressive buckling," Science Advances **7**(43), eabj3686 (2021).

[22] Y. Zhang, Z. Yan, K. Nan, D. Xiao, Y. Liu, H. Luan, H. Fu, X. Wang, Q. Yang, J. Wang, W. Ren, H. Si, F. Liu, L. Yang, H. Li, J. Wang, X. Guo, H. Luo, L. Wang, Y. Huang, and J.A. Rogers, "A mechanically driven form of Kirigami as a route to 3D mesostructures in micro/nanomembranes," Proceedings of the National Academy of Sciences **112**(38), 11757–11764 (2015).

[23] R. Su, F. Wang, and M. C. McAlpine, "3D printed microfluidics: advances in strategies, integration, and applications," Lab on a Chip **23**(5), 1279–1299 (2023).

[24] E. Prince, S. Kheiri, Y. Wang, F. Xu, J. Cruickshank, V. Topolskaia, H. Tao, E.W.K. Young, Alison.P. McGuigan, D.W. Cescon, and E. Kumacheva, "Microfluidic Arrays of Breast Tumor Spheroids for Drug Screening and Personalized Cancer Therapies," Advanced Healthcare Materials **11**(1), 2101085 (2022).

[25] S. Wiedenmann, M. Breunig, J. Merkle, C. von Toerne, T. Georgiev, M. Moussus, L. Schulte, T. Seufferlein, M. Sterr, H. Lickert, S.E. Weissinger, P. Möller, S.M. Hauck, M. Hohwieler, A. Kleger, and M. Meier, "Single-cell-resolved differentiation of human induced pluripotent stem cells into pancreatic duct-like organoids on a microwell chip," Nat Biomed Eng **5**(8), 897–913 (2021).

[26] S. Jiang, H. Zhao, W. Zhang, J. Wang, Y. Liu, Y. Cao, H. Zheng, Z. Hu, S. Wang, Y. Zhu, W. Wang, S. Cui, P.E. Lobie, L. Huang, and S. Ma, "An Automated Organoid Platform with Inter-organoid Homogeneity and Inter-patient Heterogeneity," CR Med **1**(9), (2020).

[27] Y. Wang, M. Liu, Y. Zhang, H. Liu, and L. Han, "Recent methods of droplet microfluidics and their applications in spheroids and organoids," Lab on a Chip **23**(5), 1080–1096 (2023).

[28] S. Sart, G. Ronteix, S. Jain, G. Amselem, and C.N. Baroud, "Cell Culture in Microfluidic Droplets," Chem. Rev. **122**(7), 7061–7096 (2022).

[29] A. Tourovskaia, X. Figueroa-Masot, and A. Folch, "Differentiation-on-a-chip: A microfluidic platform for long-term cell culture studies," Lab on a Chip **5**(1), 14–19 (2005).

[30] Y. Wang, L. Wang, Y. Guo, Y. Zhu, and J. Qin, "Engineering stem cell-derived 3D brain organoids in a perfusable organ-on-a-chip system," RSC Advances **8**(3), 1677–1685 (2018).

[31] S. Grebenyuk, A.R. Abdel Fattah, M. Kumar, B. Toprakhisar, G. Rustandi, A. Vananroye, I. Salmon, C. Verfaillie, M. Grillo, and A. Ranga, "Large-scale perfused tissues via synthetic 3D soft microfluidics," Nat Commun **14**(1), 193 (2023).

[32] Y. Zhu, L. Wang, H. Yu, F. Yin, Y. Wang, H. Liu, L. Jiang, and J. Qin, "In situ generation of human brain organoids on a micropillar array," Lab Chip **17**(17), 2941–2950 (2017).

[33] Y. Wang, H. Wang, P. Deng, W. Chen, Y. Guo, T. Tao, and J. Qin, "In situ differentiation and generation of functional liver organoids from human iPSCs in a 3D perfusable chip system," Lab Chip **18**(23), 3606–3616 (2018).

[34] C. Quintard, E. Tubbs, G. Jonsson, J. Jiao, J. Wang, N. Werschler, C. Laporte, A. Pitaval, T.-S. Bah, G. Pomeranz, C. Bissardon, J. Kaal, A. Leopoldi, D.A. Long, P. Blandin, J.-L. Achard, C. Battail, A. Hagelkruys, F. Navarro, Y. Fouillet, J.M. Penninger, and X. Gidrol, "A

microfluidic platform integrating functional vascularized organoids-on-chip," Nat Commun **15**(1), 1452 (2024).

[35] Y. Nashimoto, R. Mukomoto, T. Imaizumi, T. Terai, S. Shishido, K. Ino, R. Yokokawa, T. Miura, K. Onuma, M. Inoue, and H. Shiku, "Electrochemical sensing of oxygen metabolism for a three-dimensional cultured model with biomimetic vascular flow," Biosensors and Bioelectronics **219**, 114808 (2023).

[36] K. Shik Mun, K. Arora, Y. Huang, F. Yang, S. Yarlagadda, Y. Ramananda, M. Abu-El-Haija, J.J. Palermo, B.N. Appakalai, J.D. Nathan, and A.P. Naren, "Patient-derived pancreas-on-a-chip to model cystic fibrosis-related disorders," Nat Commun **10**(1), 3124 (2019).

[37] M. Kasendra, A. Tovaglieri, A. Sontheimer-Phelps, S. Jalili-Firoozinezhad, A. Bein, A. Chalkiadaki, W. Scholl, C. Zhang, H. Rickner, C.A. Richmond, H. Li, D.T. Breault, and D.E. Ingber, "Development of a primary human Small Intestine-on-a-Chip using biopsy-derived organoids," Sci Rep **8**(1), 2871 (2018).

[38] I. Salmon, S. Grebenyuk, A.R.A. Fattah, G. Rustandi, T. Pilkington, C. Verfaillie, and A. Ranga, "Engineering neurovascular organoids with 3D printed microfluidic chips," Lab on a Chip **22**(8), 1615–1629 (2022).

[39] F. Yin, X. Zhang, L. Wang, Y. Wang, Y. Zhu, Z. Li, T. Tao, W. Chen, H. Yu, and J. Qin, "HiPSC-derived multi-organoids-on-chip system for safety assessment of antidepressant drugs," Lab Chip **21**(3), 571–581 (2021).

[40] S.A.P. Rajan, J. Aleman, M. Wan, N. Pourhabibi Zarandi, G. Nzou, S. Murphy, C.E. Bishop, H. Sadri-Ardekani, T. Shupe, A. Atala, A.R. Hall, and A. Skardal, "Probing prodrug metabolism and reciprocal toxicity with an integrated and humanized multi-tissue organ-on-a-chip platform," Acta Biomaterialia **106**, 124–135 (2020).

[41] Y.S. Zhang, J. Aleman, S.R. Shin, T. Kilic, D. Kim, S.A. Mousavi Shaegh, S. Massa, R. Riahi, S. Chae, N. Hu, H. Avci, W. Zhang, A. Silvestri, A. Sanati Nezhad, A. Manbohi, F. De Ferrari, A. Polini, G. Calzone, N. Shaikh, P. Alerasool, E. Budina, J. Kang, N. Bhise, J. Ribas, A. Pourmand, A. Skardal, T. Shupe, C.E. Bishop, M.R. Dokmeci, A. Atala, and A. Khademhosseini, "Multisensor-integrated organs-on-chips platform for automated and continual in situ monitoring of organoid behaviors," Proceedings of the National Academy of Sciences **114**(12), E2293–E2302 (2017).

[42] T. Takebe, and J.M. Wells, "Organoids by design," Science **364**(6444), 956–959 (2019).

[43] E. Garreta, R.D. Kamm, S.M. Chuva de Sousa Lopes, M.A. Lancaster, R. Weiss, X. Trepat, I. Hyun, and N. Montserrat, "Rethinking organoid technology through bioengineering," Nat. Mater. **20**(2), 145–155 (2021).

[44] M. Ghosheh, A. Ehrlich, K. Ioannidis, M. Ayyash, I. Goldfracht, M. Cohen, A. Fischer, Y. Mintz, L. Gepstein, and Y. Nahmias, "Electro-metabolic coupling in multi-chambered vascularized human cardiac organoids," Nat. Biomed. Eng **7**(11), 1493–1513 (2023).

[45] M. Liu, N. Jiang, C. Qin, Y. Xue, J. Wu, Y. Qiu, Q. Yuan, C. Chen, L. Huang, L. Zhuang, and P. Wang, "Multimodal spatiotemporal monitoring of basal stem cell-derived organoids reveals progression of olfactory dysfunction in Alzheimer's disease," Biosensors and Bioelectronics **246**, 115832 (2024).

[46] Z. Lin, W. Wang, R. Liu, Q. Li, J. Lee, C. Hirschler, and J. Liu, "Cyborg organoids integrated with stretchable nanoelectronics can be functionally mapped during development," Nat Protoc, 1–32 (2025).

[47] J. Lee, W. Wang, Q. Li, Z. Lin, R. Liu, Z. Tang, J. Aoyama, R.T. Lee, X. Wang, and J. Liu, "In Situ Graphene-Seq: Spatial Transcriptomics and Chronic Electrophysiological Characterization of Tissue Microenvironments," 2025.03.25.645278 (2025).

[48] Q. Li, K. Nan, P. Le Floch, Z. Lin, H. Sheng, T.S. Blum, and J. Liu, "Cyborg Organoids: Implantation of Nanoelectronics via Organogenesis for Tissue-Wide Electrophysiology," Nano Lett. **19**(8), 5781–5789 (2019).

[49] Y. Park, C.K. Franz, H. Ryu, H. Luan, K.Y. Cotton, J.U. Kim, T.S. Chung, S. Zhao, A. Vazquez-Guardado, D.S. Yang, K. Li, R. Avila, J.K. Phillips, M.J. Quezada, H. Jang, S.S. Kwak, S.M. Won, K. Kwon, H. Jeong, A.J. Bandodkar, M. Han, H. Zhao, G.R. Osher, H. Wang, K. Lee, Y. Zhang, Y. Huang, J.D. Finan, and J.A. Rogers, "Three-dimensional, multifunctional neural interfaces for cortical spheroids and engineered assembloids," Sci. Adv. **7**(12), eabf9153 (2021).

[50] X. Yang, C. Forró, T.L. Li, Y. Miura, T.J. Zaluska, C.-T. Tsai, S. Kanton, J.P. McQueen, X. Chen, V. Mollo, F. Santoro, S.P. Pașca, and B. Cui, "Kirigami electronics for long-term electrophysiological recording of human neural organoids and assembloids," Nat Biotechnol, 1–8 (2024).

[51] H. Ryu, Y. Park, H. Luan, G. Dalgin, K. Jeffris, H.-J. Yoon, T.S. Chung, J.U. Kim, S.S. Kwak, G. Lee, H. Jeong, J. Kim, W. Bai, J. Kim, Y.H. Jung, A.K. Tryba, J.W. Song, Y. Huang, L.H. Philipson, J.D. Finan, and J.A. Rogers, "Transparent, Compliant 3D Mesostructures for Precise Evaluation of Mechanical Characteristics of Organoids," Advanced Materials **33**(25), 2100026 (2021).

[52] M. McDonald, D. Sebinger, L. Brauns, L. Gonzalez-Cano, Y. Menuchin-Lasowski, M. Mierzejewski, O.-E. Psathaki, A. Stumpf, J. Wickham, T. Rauen, H. Schöler, and P.D. Jones, "A mesh microelectrode array for non-invasive electrophysiology within neural organoids," Biosensors and Bioelectronics **228**, 115223 (2023).

[53] G.N. Angotzi, L. Giantomasi, J.F. Ribeiro, M. Crepaldi, M. Vincenzi, D. Zito, and L. Berdondini, "Integrated Micro-Devices for a Lab-in-Organoid Technology Platform: Current Status and Future Perspectives," Frontiers in Neuroscience **16**, (2022).

[54] A. Kalmykov, J.W. Reddy, E. Bedoyan, Y. Wang, R. Garg, S.K. Rastogi, D. Cohen-Karni, M. Chamanzar, and T. Cohen-Karni, "Bioelectrical interfaces with cortical spheroids in three-dimensions," J. Neural Eng. **18**(5), 055005 (2021).

[55] B. Nasr, R. Chatterton, J.H.M. Yong, P. Jamshidi, G.M. D'Abaco, A.R. Bjorksten, O. Kavehei, G. Chana, M. Dottori, and E. Skafidas, "Self-Organized Nanostructure Modified Microelectrode for Sensitive Electrochemical Glutamate Detection in Stem Cells-Derived Brain Organoids," Biosensors (Basel) **8**(1), 14 (2018).

[56] M. Devarasetty, S. Forsythe, T.D. Shupe, S. Soker, C.E. Bishop, A. Atala, and A. Skardal, "Optical Tracking and Digital Quantification of Beating Behavior in Bioengineered Human Cardiac Organoids," Biosensors **7**(3), 24 (2017).

[57] Q. Lyu, S. Gong, J.G. Lees, J. Yin, L.W. Yap, A.M. Kong, Q. Shi, R. Fu, Q. Zhu, A. Dyer, J.M. Dyson, S.Y. Lim, and W. Cheng, "A soft and ultrasensitive force sensing diaphragm for probing cardiac organoids instantaneously and wirelessly," Nat Commun **13**(1), 7259 (2022).

[58] G.N. Angotzi, M. Crepaldi, A. Lecomte, L. Giantomasi, S. Rancati, D.D. Tonelli, and L. Berdondini, "A μRadio CMOS Device for Real-Time In-Tissue Monitoring of Human Organoids," in *2018 IEEE Biomedical Circuits and Systems Conference (BioCAS)*, (2018), pp. 1–4.

[59] O. Phouphetlinthong, E. Partiot, C. Bernou, A. Sebban, R. Gaudin, and B. Charlot, "Protruding cantilever microelectrode array to monitor the inner electrical activity of cerebral organoids," Lab on a Chip **23**(16), 3603–3614 (2023).

[60] M. Kim, J.C. Hwang, S. Min, Y.-G. Park, S. Kim, E. Kim, H. Seo, W.G. Chung, J. Lee, S.-W. Cho, and J.-U. Park, "Multimodal Characterization of Cardiac Organoids Using Integrations of Pressure-Sensitive Transistor Arrays with Three-Dimensional Liquid Metal Electrodes," Nano Lett. **22**(19), 7892–7901 (2022).

[61] C. Cassel de Camps, S. Aslani, N. Stylianesis, H. Nami, N.-V. Mohamed, T.M. Durcan, and C. Moraes, "Hydrogel Mechanics Influence the Growth and Development of Embedded Brain Organoids," ACS Appl. Bio Mater. **5**(1), 214–224 (2022).

[62] B. Buchmann, P. Fernández, and A.R. Bausch, "The role of nonlinear mechanical properties of biomimetic hydrogels for organoid growth," Biophysics Reviews **2**(2), 021401 (2021).

[63] M.S. Yousafzai, and J.A. Hammer, "Using Biosensors to Study Organoids, Spheroids and Organs-on-a-Chip: A Mechanobiology Perspective," Biosensors **13**(10), 905 (2023).

[64] S. Alhudaithy, and K. Hoshino, "Biocompatible micro tweezers for 3D hydrogel organoid array mechanical characterization," PLoS One **17**(1), e0262950 (2022).

[65] J. Yin, J.G. Lees, S. Gong, J.T. Nguyen, R.J. Phang, Q. Shi, Y. Huang, A.M. Kong, J.M. Dyson, S.Y. Lim, and W. Cheng, "Real-time electro-mechanical profiling of dynamically beating human cardiac organoids by coupling resistive skins with microelectrode arrays," Biosensors and Bioelectronics **267**, 116752 (2025).

[66] K.F. Reardon, "Practical monitoring technologies for cells and substrates in biomanufacturing," Current Opinion in Biotechnology **71**, 225–230 (2021).

[67] D.J. Carvalho, A.M. Kip, A. Tegel, M. Stich, C. Krause, M. Romitti, C. Branca, B. Verhoeven, S. Costagliola, L. Moroni, and S. Giselbrecht, "A Modular Microfluidic Organoid Platform Using LEGO-Like Bricks," Advanced Healthcare Materials **n/a**(n/a), 2303444 (2024).

[68] D. Bavli, S. Prill, E. Ezra, G. Levy, M. Cohen, M. Vinken, J. Vanfleteren, M. Jaeger, and Y. Nahmias, "Real-time monitoring of metabolic function in liver-onchip microdevices tracks the dynamics of Mitochondrial dysfunction," Proceedings of the National Academy of Sciences of the United States of America **113**(16), E2231–E2240 (2016).

[69] E.L. Brooks, K.K. Hussain, K. Kotecha, A. Abdalla, and B.A. Patel, "Three-Dimensional-Printed Electrochemical Multiwell Plates for Monitoring Food Intolerance from Intestinal Organoids," ACS Sens. **8**(2), 712–720 (2023).

[70] R. Riahi, S.A.M. Shaegh, M. Ghaderi, Y.S. Zhang, S.R. Shin, J. Aleman, S. Massa, D. Kim, M.R. Dokmeci, and A. Khademhosseini, "Automated microfluidic platform of bead-based electrochemical immunosensor integrated with bioreactor for continual monitoring of cell secreted biomarkers," Sci Rep **6**(1), 24598 (2016).

[71] S. Spitz, S. Bolognin, K. Brandauer, J. Füßl, P. Schuller, S. Schobesberger, C. Jordan, B. Schädl, J. Grillari, H.D. Wanzenboeck, T. Mayr, M. Harasek, J.C. Schwamborn, and P. Ertl, "Development of a multi-sensor integrated midbrain organoid-on-a-chip platform for studying Parkinson's disease," 2022.08.19.504522 (2022).

[72] S.R. Shin, Y.S. Zhang, D.-J. Kim, A. Manbohi, H. Avci, A. Silvestri, J. Aleman, N. Hu, T. Kilic, W. Keung, M. Righi, P. Assawes, H.A. Alhadrami, R.A. Li, M.R. Dokmeci, and A. Khademhosseini, "Aptamer-Based Microfluidic Electrochemical Biosensor for Monitoring Cell-Secreted Trace Cardiac Biomarkers," Anal. Chem. **88**(20), 10019–10027 (2016).

73 M. Barroso, M.G. Monaghan, R. Niesner, and R.I. Dmitriev, "Probing organoid metabolism using fluorescence lifetime imaging microscopy (FLIM): The next frontier of drug discovery and disease understanding," Advanced Drug Delivery Reviews **201**, 115081 (2023).

74 W. H. Skinner, N. Robinson, G. R. Hardisty, H. Fleming, A. Geddis, M. Bradley, R. D. Gray, and C. J. Campbell, "SERS microsensors for pH measurements in the lumen and ECM of stem cell derived human airway organoids," Chemical Communications **59**(22), 3249–3252 (2023).

75 J. Dornhof, J. Kieninger, H. Muralidharan, J. Maurer, G.A. Urban, and A. Weltin, "OXYGEN AND LACTATE MONITORING IN 3D BREAST CANCER ORGANOID CULTURE WITH SENSOR-INTEGRATED MICROFLUIDIC PLATFORM," in *2021 21st International Conference on Solid-State Sensors, Actuators and Microsystems (Transducers)*, (2021), pp. 703–706.

76 Y. Teng, Z. Zhao, F. Tasnim, X. Huang, and H. Yu, "A scalable and sensitive steatosis chip with long-term perfusion of in situ differentiated HepaRG organoids," Biomaterials **275**, 120904 (2021).

77 K. Zhang, J. Xi, Y. Wang, J. Xue, B. Li, Z. Huang, Z. Zheng, N. Liang, and Z. Wei, "A Microfluidic Chip-Based Automated System for Whole-Course Monitoring the Drug Responses of Organoids," Anal. Chem. **96**(24), 10092–10101 (2024).

78 Y. Wang, L. Wang, Y. Zhu, and J. Qin, "Human brain organoid-on-a-chip to model prenatal nicotine exposure," Lab Chip **18**(6), 851–860 (2018).

79 D. Sun, W. Gao, H. Hu, and S. Zhou, "Why 90% of clinical drug development fails and how to improve it?," Acta Pharmaceutica Sinica B **12**(7), 3049–3062 (2022).

80 T.M. Cannon, A.T. Shah, and M.C. Skala, "Autofluorescence imaging captures heterogeneous drug response differences between 2D and 3D breast cancer cultures," Biomed. Opt. Express, BOE **8**(3), 1911–1925 (2017).

81 S. Kim, S. Choung, R.X. Sun, N. Ung, N. Hashemi, E.J. Fong, R. Lau, E. Spiller, J. Gasho, J. Foo, and S.M. Mumenthaler, "Comparison of Cell and Organoid-Level Analysis of Patient-Derived 3D Organoids to Evaluate Tumor Cell Growth Dynamics and Drug Response," SLAS DISCOVERY: Advancing the Science of Drug Discovery **25**(7), 744–754 (2020).

82 A. Brooks, X. Liang, Y. Zhang, C.-X. Zhao, M.S. Roberts, H. Wang, L. Zhang, and D.H.G. Crawford, "Liver organoid as a 3D in vitro model for drug validation and toxicity assessment," Pharmacological Research **169**, 105608 (2021).

83 P.W. Nagle, J.Th.M. Plukker, C.T. Muijs, P. van Luijk, and R.P. Coppes, "Patient-derived tumor organoids for prediction of cancer treatment response," Seminars in Cancer Biology **53**, 258–264 (2018).

84 J. Kondo, and M. Inoue, "Application of Cancer Organoid Model for Drug Screening and Personalized Therapy," Cells **8**(5), 470 (2019).

85 L. Li, H. Knutsdottir, K. Hui, M.J. Weiss, J. He, B. Philosophe, A.M. Cameron, C.L. Wolfgang, T.M. Pawlik, G. Ghiaur, A.J. Ewald, E. Mezey, J.S. Bader, and F.M. Selaru, "Human primary liver cancer organoids reveal intratumor and interpatient drug response heterogeneity," JCI Insight **4**(2), e121490 (n.d.).

86 A. Skardal, M. Devarasetty, C. Rodman, A. Atala, and S. Soker, "Liver-Tumor Hybrid Organoids for Modeling Tumor Growth and Drug Response In Vitro," Ann Biomed Eng **43**(10), 2361–2373 (2015).

[87] P.-O. Frappart, and T.G. Hofmann, "Pancreatic Ductal Adenocarcinoma (PDAC) Organoids: The Shining Light at the End of the Tunnel for Drug Response Prediction and Personalized Medicine," Cancers **12**(10), 2750 (2020).

[88] P.-O. Frappart, K. Walter, J. Gout, A.K. Beutel, M. Morawe, F. Arnold, M. Breunig, T.F. Barth, R. Marienfeld, L. Schulte, T. Ettrich, T. Hackert, M. Svinarenko, R. Rösler, S. Wiese, H. Wiese, L. Perkhofer, M. Müller, A. Lechel, B. Sainz, P.C. Hermann, T. Seufferlein, and A. Kleger, "Pancreatic cancer-derived organoids – a disease modeling tool to predict drug response," United European Gastroenterology Journal **8**(5), 594–606 (2020).

[89] A. Hennig, F. Baenke, A. Klimova, S. Drukewitz, B. Jahnke, S. Brückmann, R. Secci, C. Winter, T. Schmäche, T. Seidlitz, J.-P. Bereuter, H. Polster, L. Eckhardt, S.A. Schneider, S. Brückner, R. Schmelz, J. Babatz, C. Kahlert, M. Distler, J. Hampe, M. Reichert, S. Zeißig, G. Folprecht, J. Weitz, D. Aust, T. Welsch, and D.E. Stange, "Detecting drug resistance in pancreatic cancer organoids guides optimized chemotherapy treatment," The Journal of Pathology **257**(5), 607–619 (2022).

[90] E. Campaner, A. Zannini, M. Santorsola, D. Bonazza, C. Bottin, V. Cancila, C. Tripodo, M. Bortul, F. Zanconati, S. Schoeftner, and G. Del Sal, "Breast Cancer Organoids Model Patient-Specific Response to Drug Treatment," Cancers **12**(12), 3869 (2020).

[91] C.G. Hubert, M. Rivera, L.C. Spangler, Q. Wu, S.C. Mack, B.C. Prager, M. Couce, R.E. McLendon, A.E. Sloan, and J.N. Rich, "A Three-Dimensional Organoid Culture System Derived from Human Glioblastomas Recapitulates the Hypoxic Gradients and Cancer Stem Cell Heterogeneity of Tumors Found In Vivo," Cancer Research **76**(8), 2465–2477 (2016).

[92] S. Frenz-Wiessner, S.D. Fairley, M. Buser, I. Goek, K. Salewskij, G. Jonsson, D. Illig, B. zu Putlitz, D. Petersheim, Y. Li, P.-H. Chen, M. Kalauz, R. Conca, M. Sterr, J. Geuder, Y. Mizoguchi, R.T.A. Megens, M.I. Linder, D. Kotlarz, M. Rudelius, J.M. Penninger, C. Marr, and C. Klein, "Generation of complex bone marrow organoids from human induced pluripotent stem cells," Nat Methods **21**(5), 868–881 (2024).

[93] D.S. Park, T. Kozaki, S.K. Tiwari, M. Moreira, A. Khalilnezhad, F. Torta, N. Olivié, C.H. Thiam, O. Liani, A. Silvin, W.W. Phoo, L. Gao, A. Triebl, W.K. Tham, L. Gonçalves, W.T. Kong, S. Raman, X.M. Zhang, G. Dunsmore, C.A. Dutertre, S. Lee, J.M. Ong, A. Balachander, S. Khalilnezhad, J. Lum, K. Duan, Z.M. Lim, L. Tan, I. Low, K.H. Utami, X.Y. Yeo, S. Di Tommaso, J.-W. Dupuy, B. Varga, R.T. Karadottir, M.C. Madathummal, I. Bonne, B. Malleret, Z.Y. Binte, N. Wei Da, Y. Tan, W.J. Wong, J. Zhang, J. Chen, R.M. Sobota, S.W. Howland, L.G. Ng, F. Saltel, D. Castel, J. Grill, V. Minard, S. Albani, J.K.Y. Chan, M.S. Thion, S.Y. Jung, M.R. Wenk, M.A. Pouladi, C. Pasqualini, V. Angeli, O.N.F. Cexus, and F. Ginhoux, "iPS-cell-derived microglia promote brain organoid maturation via cholesterol transfer," Nature **623**(7986), 397–405 (2023).

[94] E. Karzbrun, A. Kshirsagar, S.R. Cohen, J.H. Hanna, and O. Reiner, "Human brain organoids on a chip reveal the physics of folding," Nature Phys **14**(5), 515–522 (2018).

[95] T. Tao, Y. Wang, W. Chen, Z. Li, W. Su, Y. Guo, P. Deng, and J. Qin, "Engineering human islet organoids from iPSCs using an organ-on-chip platform," Lab on a Chip **19**(6), 948–958 (2019).

[96] M. Edde, G. Leroux, E. Altena, and S. Chanraud, "Functional brain connectivity changes across the human life span: From fetal development to old age," Journal of Neuroscience Research **99**(1), 236–262 (2021).

[97] K. Grenier, J. Kao, and P. Diamandis, "Three-dimensional modeling of human neurodegeneration: brain organoids coming of age," Mol Psychiatry **25**(2), 254–274 (2020).

[98] S.J. Mun, Y.-H. Hong, H.-S. Ahn, J.-S. Ryu, K.-S. Chung, and M.J. Son, "Long-Term Expansion of Functional Human Pluripotent Stem Cell-Derived Hepatic Organoids," Int J Stem Cells **13**(2), 279–286 (2020).